\documentclass[fleqn,usenatbib]{mnras}

\usepackage{newtxtext,newtxmath}
\usepackage{amsmath}
\usepackage[T1]{fontenc}
\usepackage{soul}

\DeclareRobustCommand{\VAN}[3]{#2}
\let\VANthebibliography\thebibliography
\def\thebibliography{\DeclareRobustCommand{\VAN}[3]{##3}\VANthebibliography}

\usepackage{graphicx}	
\usepackage{amsmath}	

\title[
]{Population III star formation near high-redshift active galactic nuclei}

\author[Fisk et al.]{
Ethan M. Fisk$^{1,2}$\thanks{E-mail: emf@lanl.gov},
Madeline A. Marshall$^1$,
Phoebe R. Upton Sanderbeck$^{1}$,
Jarrett L. Johnson$^{1}$
\\
$^{1}$Los Alamos National Laboratory, Los Alamos, NM 87545, USA\\
$^{2}$Department of Physics, Applied Physics, and Astronomy, Rensselaer Polytechnic Institute, Troy, NY 12180, USA\\
}

\date{Accepted 2026 March 18. Received 2026 January 12; in original form 2025 August 14 }

\pubyear{\the\year{}}

\begin{document}
\label{firstpage}
\pagerange{\pageref{firstpage}--\pageref{lastpage}}
\maketitle

\begin{abstract}
Using cosmological radiation-hydrodynamical simulations, we study the effect of accreting supermassive black holes (SMBHs) on nearby dark-matter (DM) haloes in the very early universe. We find that an SMBH with a spectral energy distribution (SED) extending from the near-ultraviolet to hard X-rays, can produce a radiation background sufficient to delay gravitational collapse in surrounding DM haloes until up to $10^7$ M$_\odot$ of zero-metallicity gas is available for the formation of Population III (Pop III) stars or  direct-collapse black holes (DCBHs). We model three scenarios, corresponding to an SMBH located at physical distances of 10, 100, and 1000 kpc from the Pop III host DM halo. Using these three scenarios, we use the SED to compute self-consistent photoionization, photoheating, and photodissociation rates. 
We include the effects of Compton scattering 
and gas self-shielding. The X-ray portion of the spectrum maintains an elevated free-electron fraction as the gas collapses to high density. This stimulates H$_2$ formation, allowing the gas to cool further while counteracting the dissociation of H$_2$ by Lyman-Werner radiation. As a result, a large cluster of Pop III stars is expected to form, except in the case with the most intense radiation in which a DCBH may instead form. Our simulated Pop III clusters have comparable He\,{\large \sc ii} $\lambda$1640 luminosities to the recently discovered Pop III host candidate near GN-z11, observed by the \textit{James Webb Space Telescope}. In two of the scenarios we consider, the resulting clusters could be detectable using the telescope's NIRSpec instrument out to $z$ $\sim$ 15.
\end{abstract}




\section{Introduction}
Population III (Pop III) stars, the first generation of stars in the universe formed from primordial gas, have yet to be definitively discovered. They are an exciting target for the \textit{James Webb Space Telescope} (\textit{JWST}) \citep{Gardner2006,Gardner2023,McElwain2023}, which contains several infrared-detecting instruments that allow the telescope to observe high-redshift phenomena. Recently, a possible He\,{\large \sc ii} $\lambda$1640 emission line, an expected spectral signature of Pop III  stars, was observed in the vicinity of the galaxy GN-z11  at $z = 10.6$ \citep{Maiolino2024}. Another galaxy, LAP1-B, observed at $z = 6.6$, has very low metallicity and a hard ionizing spectrum \citep{nakajima2025} that aligns with theoretical predictions for Pop III stars. \citet{Fujimoto2025} presented an upper limit on the luminosity function of Pop III galaxies at $z = 6-7$, and \citet{Hsiao2025} observed 12 very low-metallicity galaxy candidates at $z \sim 5-7$, two of which have metallicities $\rm Z < 10^{-2} \rm\ Z_{\odot}$. An even lower upper limit of $\rm Z < 10^{-3} \rm\ Z_{\odot}$ has been reported for a gravitationally-lensed galaxy at $z = 5.8$ \citep{Morishita2025}. Low-metallicity galaxies have also been observed at lower redshifts. \citet{mondal2025} observed He\,{\large \sc ii} $\lambda$1640 emission at $z = 2.98$, possibly powered by small pockets of Pop III stars in a partially metal-enriched galaxy. \citet{cai2025metalfreegalaxyz} also discovered the galaxy MPG-CR3 at $z = 3.19$, with a metallicity $\rm Z < 8 \times 10^{-3}\ \rm Z_{\odot}$ and an estimated stellar mass of $6.1 \times 10^5\ \rm M_{\odot}$.

Previous work shows that the characteristics of Pop III clusters would be strongly shaped by their radiative environment. An H$_2$-dissociating Lyman-Werner (LW) background can prevent the gas from cooling, increasing the initial mass function (IMF) of the stars formed, and with a high enough flux inducing the formation of a direct-collapse black hole \citep[DCBH;][]{Latif2014}. Other work also showed that an ionizing ultraviolet (UV) background could delay Pop III formation until lower redshifts \citep{Visbal2017,Kulkarni2019}. \citet{Park2021} showed that a much weaker X-ray background could elevate the free-electron fraction in the collapsing gas, catalysing H$_2$ formation, which boosts Pop III star formation in low-mass dark matter (DM) haloes.

Past work has also studied Pop III formation in the most extreme radiative environments.  \citet{Johnson2019} showed, using a simplified one-zone model, that a monochromatic 1000 eV flux from the quasar of a nearby supermassive black hole (SMBH) could also induce the formation of a DCBH within 100 kpc of the host DM halo. \citet{Prole2024}
also showed that a DM halo merger can increase the gas temperature of the resultant halo through dynamical heating, increasing the initial mass function (IMF) of the objects that form without a radiation background. A related scenario, although not involving Pop III stars, was observed by \citet{Balashev2025}, which presented observations of two merging galaxies (one of which contains a powerful quasar) at $z = 2.7$. In the quasar host’s companion, compact, high-density gas clouds were found, indicating that UV from the quasar dissociated most of the molecular gas, leaving only high-density regions that were shielded from the quasar, a scenario which may also explain the inferred decline in recent star formation near a bright quasar at $z = 6.3$ reported by \citet{Zhu2025}. Similarly, \citet{Suzuki2025} showed that Lyman-$\alpha$ emitting and continuum-selected galaxies have lower densities when they are near quasars, and \citet{Durovcikova2025} observed a Lyman-$\alpha$ emitter candidate at a projected distance of 29 kpc from the quasar PSO J158-14. 

Here we extend previous work by running 3D radiation-hydrodynamical simulations to model the radiative feedback of an active galactic nucleus (AGN) on the formation of Pop III stars in neighbouring regions. We use a standard composite quasar SED \citep{HM2012} to model the entire spectrum, computing self-consistent photodissociation, photoionization, and photoheating rates. We also account for Compton scattering and inverse Compton scattering self-consistently with this spectrum. In Section \ref{Sec:methods} we discuss our simulations and in Section \ref{Sec:results} we present the results. In Section \ref{Sec:discussion} we determine the specific objects that would form after gravitational collapse in each scenario, and their detectability with \textit{JWST}. Finally, in Section \ref{Sec:conclusion} we summarize our findings. We assume a $\Lambda \rm CDM$ cosmology with the following parameters from \citet{planck}: 
$\Omega_{\rm M}$ = 0.309, 
$\Omega_{\Lambda}$ =
0.691, 
$\Omega_{\rm b}$ = 0.0486, h = 0.677, $\sigma_8$= 0.816, and $n_s$ = 0.968.

\section{Simulations and Methods}

\label{Sec:methods}
\subsection{Simulation Setup}
Using {\Large\sc enzo}, \citep{Bryan2014}, a cosmological adaptive mesh refinement (AMR) radiation hydrodynamics code, we initialized a periodic box at $z = 200$ with a comoving size of $(10  \text{ Mpc/h})^3$. Initial conditions were generated with {\Large\sc music} \citep{HahnAbel}. First, we ran a  simulation with resolution $128^3$ (128 cells per side) and no AMR. We tracked the largest DM halo using the halo finder in \texttt{yt} \citep{yt}, then restarted the simulation at $z = 200$ centred on this largest halo. By $z=12$, this halo reached a mass of  ${\sim} 2.5 \times 10^9$ M$_\odot$. During the second run, we tracked the evolution of 12 species (H, H$^+$, He, He$^+$, He$^{++}$, e$^-$, H$_2$, H$_2^+$, H$^-$, D, D$^+$, and HD).

The impact of the AGN radiation on the primordial gas was modelled using three different scenarios (described in Section \ref{sec:quasar_background}) as an isotropic radiation background. To avoid discontinuities in the simulation, we gradually ramped up the intensity of the background using a logistic function. The radiation backgrounds reached full intensity at $z = 37$, in order to ionize the entire box before gravitational collapse. \citet{Visbal2017} showed that the time at which the radiation background is turned on has no significant effect on the Pop III star clusters that form, as long as the background can completely ionize the intergalactic medium. 

In our second set of runs with the radiation background turned on (with the same box size as the first run), the base resolution remained at $128^3$  with the inner 20\% of the box being set to an effective resolution of $512^3$. AMR was also turned on in the central region based on the local Jeans length (resolved by 16 cells across), and a 1.5 overdensity criterion for both baryons and dark matter. At the highest densities, the simulations had up to 18 levels of refinement from the base grid, allowing a maximum physical resolution of 0.11 comoving pc. The simulations were stopped upon reaching a threshold gas density of $10^{-18} \rm\ g\ cm^{-3}$, after which an unresolvable runaway collapse to stellar densities would occur.

\begin{figure}
	\includegraphics[width=\columnwidth]{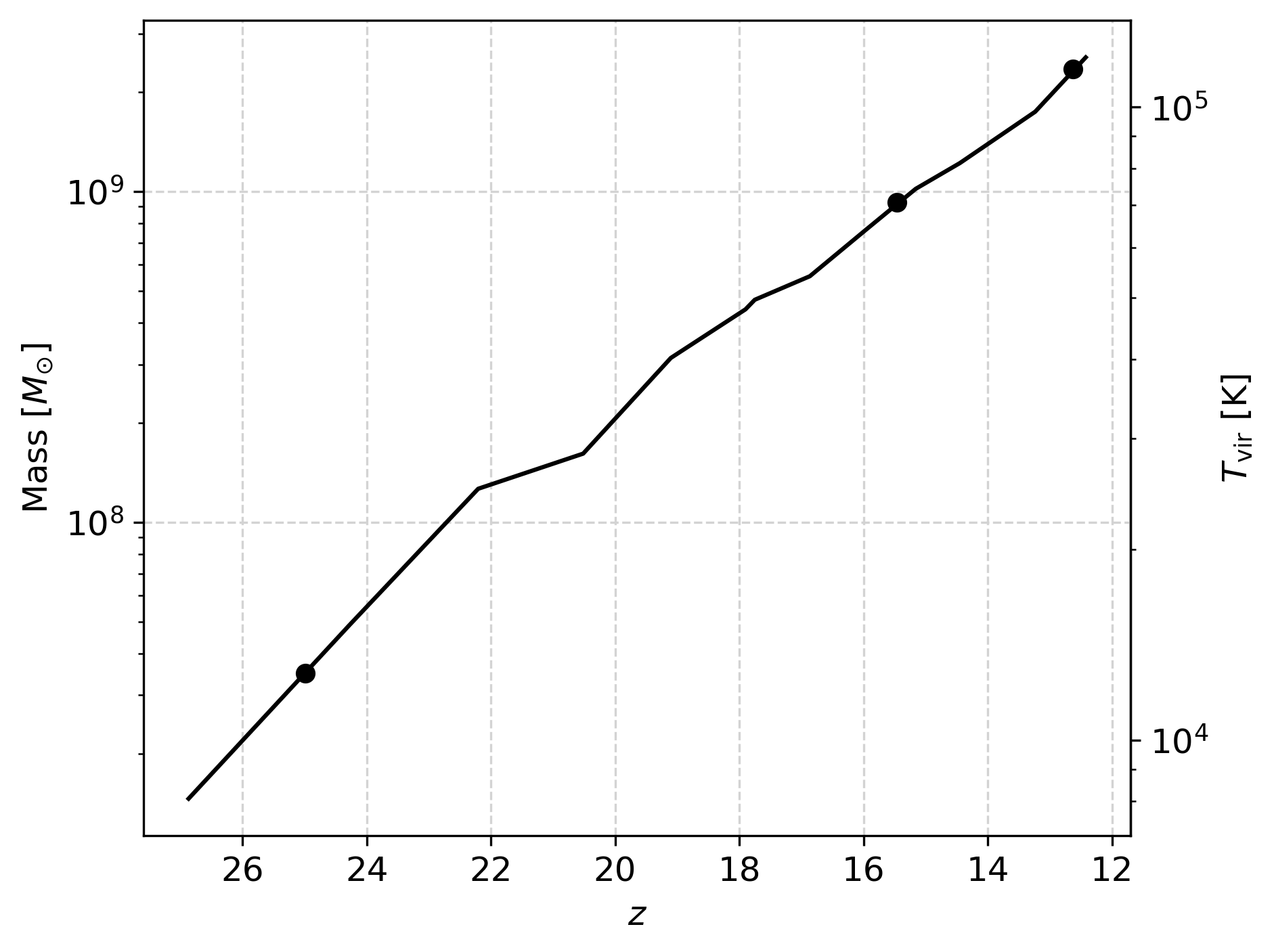}
    \caption{The virial mass of the growing DM halo (used in all three scenarios) is plotted against redshift and virial temperature. The redshifts at which runaway gravitational collapse happens (z = 24.99, 15.46, 12.63) for scenarios A, B, and C, respectively, are labelled. }
    \label{fig:virial}
\end{figure}
\subsection{Quasar Background}
\label{sec:quasar_background}
To compute our quasar background, we assume an SMBH with bolometric luminosity of $1.26 \times 10^{47}$ erg s$^{-1}$ ($L_{\rm bol}$), corresponding to an SMBH of $10^9$ M$_\odot$ accreting at the Eddington limit. 
 We simulate three different scenarios. In Scenario A, we adopt a uniform background radiation field with a flux corresponding to that which would be produced by the AGN at a distance $r = 1000$ kpc away from our DM halo. In Scenario B, the radiative flux corresponds to that produced at a distance of $r = 100$ kpc away, and Scenario C corresponds to a distance of $r = 10$ kpc. We use the quasar spectrum at $z = 0$ from \citet{HM2012}, which extends from $3-10^6\ \rm eV$, to compute a piecewise power law fit, which is normalized to the black hole mass and distance. 

 We model the three backgrounds as a set of isotropic photoionization, photoheating, photodissociation, and Compton scattering rates calculated using our chosen spectrum. Isotropic photoionization rates are calculated (in units of s$^{-1}$) using the following,
\begin{equation}
\gamma = \frac{L_{\rm bol}}{4\pi r^2} \int_{\nu_0}^{\infty} \frac{\sigma_{\nu}f_{\nu}d\nu}{h\nu},
\end{equation}
where $h$ is Planck's constant, $\sigma_{\nu}$ is the photoionization cross section of each respective species as a function of frequency, $f_\nu$ is the fraction of photons at a given frequency (where the total is normalized to one), and $\nu_0$ is the minimum ionization frequency of each respective species.
The corresponding photoheating rates are calculated as
\begin{equation}
\Gamma = \frac{L_{\rm bol}}{4\pi r^2} \int_{\nu_0}^{\infty} \frac{\sigma_{\rm \nu}f_{\rm \nu}(\nu -\nu_0)d\nu}{\nu}.
\end{equation}
Photoionization cross sections for the two hydrogenic species (H and He\,{\large \sc ii}) were taken from \citet{Osterbrock2006}, and cross sections for HeI were taken from \citet{Verner1996}. The photodissociating LW flux was calculated by normalizing the fraction of photons in the $11.2-13.6 \rm\ eV$ range to the bolometric luminosity of the SMBH. Appendix A lists the specific rates used in the simulations, based on these calculations. 

We also include Compton heating, based on \citet{RybickiLightman}, where the energy deposited to an electron in one Compton scattering is
\begin{equation} 
\Delta E = \frac{\langle h\nu \rangle}{m_{\rm e}c^2}(\langle h\nu \rangle - 4kT),
\end{equation}
where $\langle h\nu \rangle$ is the mean photon energy for our spectrum (26.95 eV), $m_{\rm e}$ is the mass of the electron, $c$ is the vacuum light speed, $k$ is the Boltzmann constant and $T$ is temperature. The rate of energy deposited per unit volume is then
\begin{equation}
\frac{dE}{dt} =\frac{\langle h\nu \rangle}{m_{\rm e}c^2}(\langle h\nu \rangle - 4kT) \sigma_{\rm T} n_{\rm e} F_\gamma,
\end{equation}
where $F_\gamma = L_{\rm bol}/(4\pi r^2 \langle h\nu \rangle)$, $\sigma_{\rm t}$ is the Thomson cross section, and $n_{\rm e}$ is the number density of free electrons.

We include shielding for photoionizations and photoheating, using a default {\Large\sc enzo} setting, although the average photoionization cross sections used for the shielding are recalculated based on the quasar spectrum \citep{HM2012}. We also include shielding for H$_2$ dissociations, based on \citet{Wolcott-Green}:
\begin{align}
f_{\mathrm{shield}}(N_{\mathrm{H}_2}, T) =\ 
&\frac{0.965}{\left( 1 + \dfrac{x}{b_5} \right)^{\alpha(n, T)}} 
\notag \\+ \frac{0.035}{(1 + x)^{0.5}} 
&\times \exp\left[ -8.5 \times 10^{-4} \, (1 + x)^{0.5} \right],
\end{align}
where $x \equiv N_{\rm H_2}/5 \times10^{14} \mathrm{cm}^{-2}$, $b_5\equiv b/10^5 \mathrm{cm\ s}^{-1}$, $\alpha$ = 1.1, $N_{\rm H_2}$ is the H$_2$ number density, and $b$ is the Doppler broadening parameter $\sqrt{\frac{2kT}{m}}$, where m is the mass of an H$_2$ molecule. 

\section{Results} 
\label{Sec:results}

\begin{figure*}
	\includegraphics[width=\textwidth]{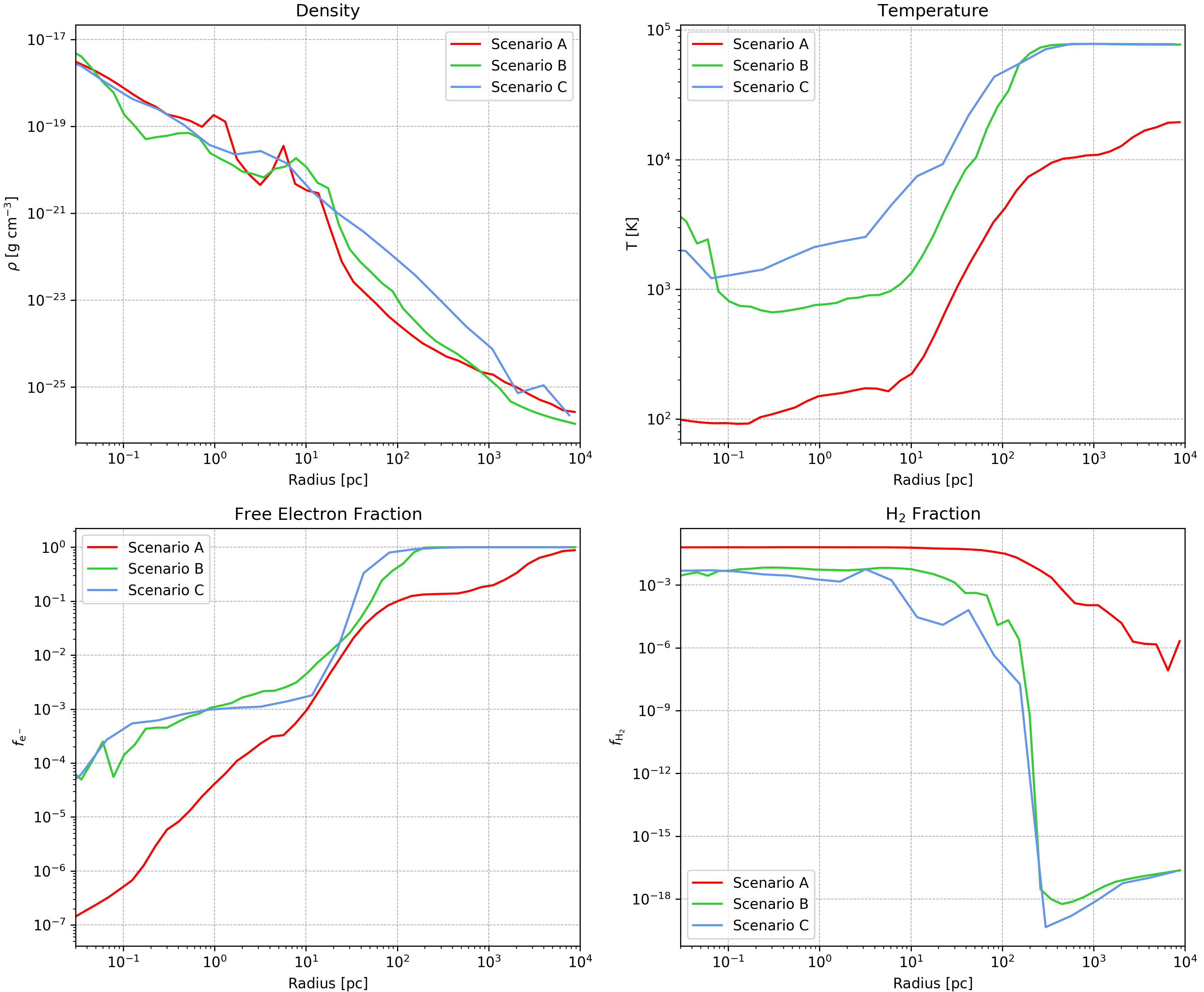}
    \caption{Radial profiles of temperature, density, H$_2$ fraction, and free-electron fraction, for each scenario, centred on the highest density cell at the moment the highest refinement level is reached (the stopping point for the simulations, after which a rapid runaway gravitational collapse would occur). The free-electron fraction stays elevated as the gas collapses to higher densities, catalysing H$_2$ formation. The LW flux counteracts this effect. }
    \label{fig:radial}
\end{figure*}
\begin{figure*}
	\includegraphics[width=\textwidth]{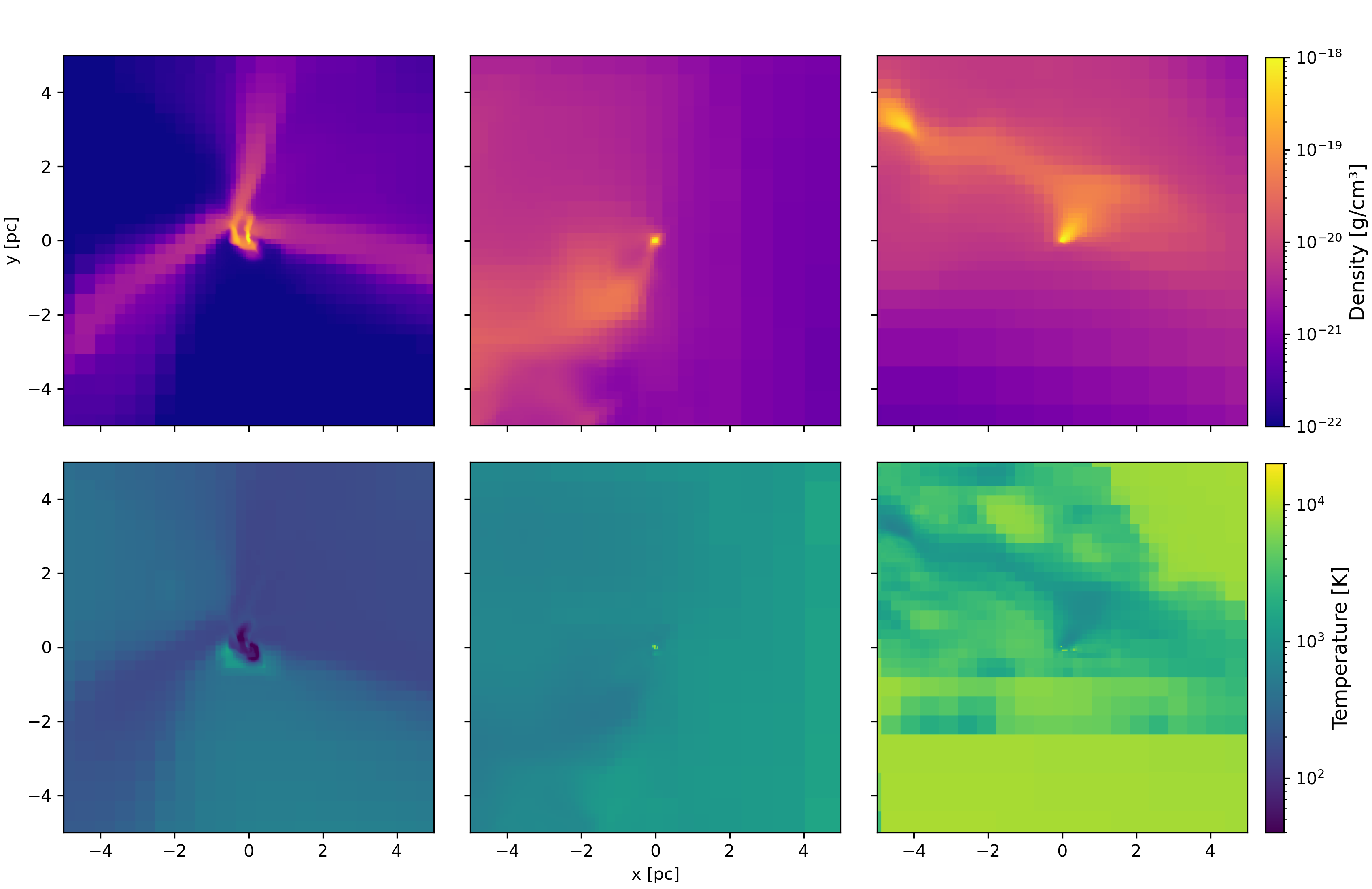}
    \caption{Density and temperature slices at the moment the highest level of refinement is reached for the lowest ({\it left}), intermediate ({\it middle}) and highest ({\it right}) background radiation fields. In Scenario A (lowest background), the gas cools to the temperature of the CMB because of HD cooling. With the two higher backgrounds (Scenarios B and C), H$_2$ formation is limited by the LW flux. }
    \label{fig:slice}
\end{figure*}

\begin{figure}
	\includegraphics[width=\columnwidth]{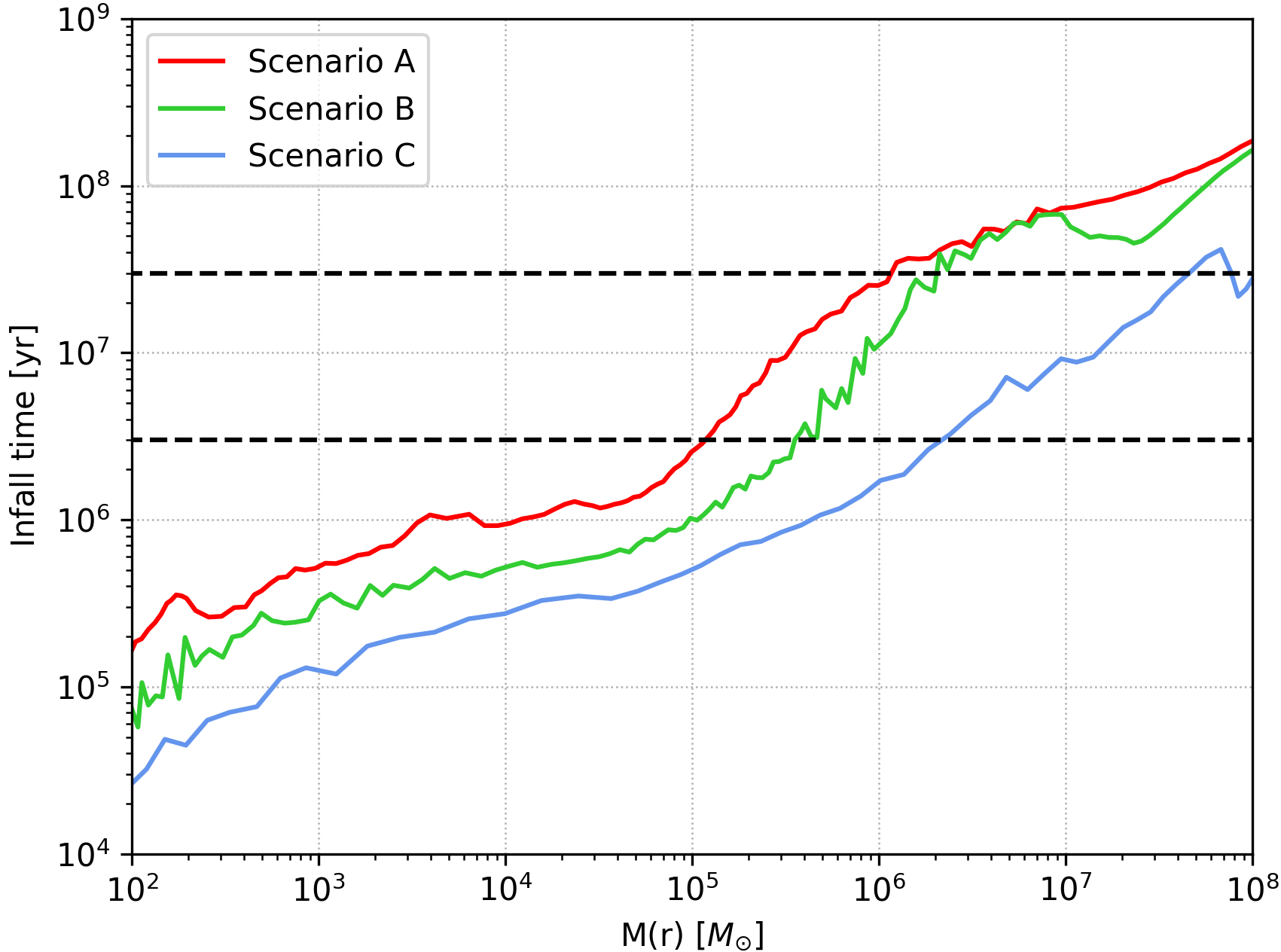}
    \caption{Infall time vs. mass enclosed, for each of the three backgrounds at the moment of collapse, using a similar methodology to \citet{Visbal2017}, where infall time is radial distance divided by spherically averaged radial velocity. Cells which had a negative radial velocity relative to the highest density cell were excluded from the calculation, because they correspond to outflowing gas that would not be accreted. The dashed lines represent 3 million and 30 million year time-scales, representing the lifetime of the most massive stars, and the maximum time limit from \citet{sarmento2025}, respectively.  }
    \label{fig:enclosed}
\end{figure}

\begin{figure}
	\includegraphics[width=\columnwidth]{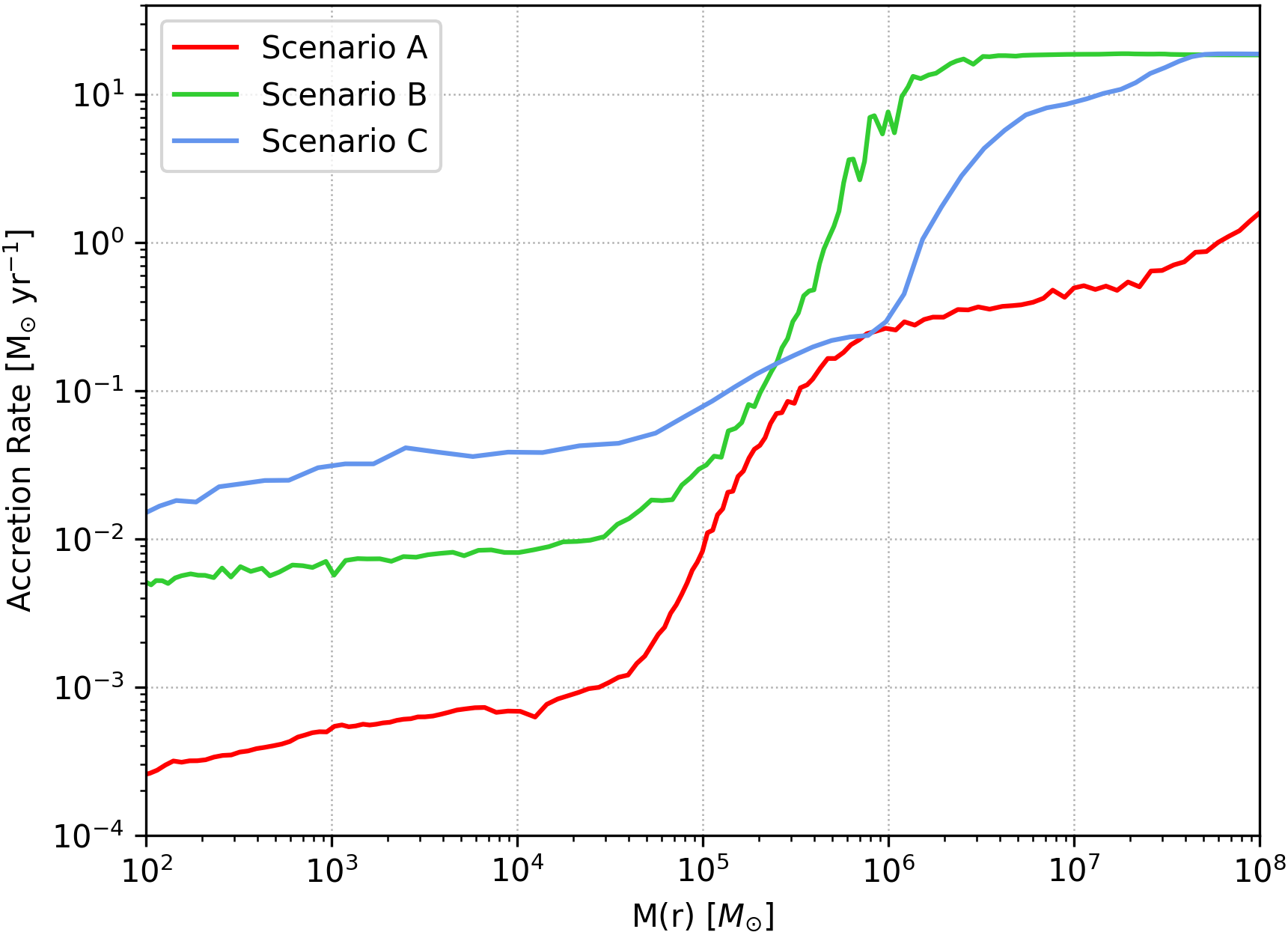}
    \caption{Accretion rate, approximated using the relation $\dot M \sim c_s^3/G$, is plotted as a function of mass enclosed. Only in Scenario C are there sustained accretion rates high enough to form a DCBH.}
    \label{fig:accretion}
\end{figure}
Here we present the results of the three simulations of the AGN at 1000, 100, and 10 kpc, which we refer to as Scenarios A, B, and C, respectively. Figure \ref{fig:virial} shows the build-up of dark matter mass in the simulated halo as a function of redshift. Runaway gravitational collapse of the primordial gas occurred at $z$ = 24.99, 15.46, and 12.63, for Scenarios A, B, and C, respectively. Our DM halo grew to a mass of $2.23\times 10^9$ M$_\odot$ at the time of runaway gravitational collapse in Scenario C. In Scenarios B and C, Compton scattering from the quasar background was powerful enough to keep the free electrons and photons in thermal equilibrium, meaning that the gas stayed ionized, preventing it from cooling via atomic electron transitions, and keeping it at a temperature of $ 78,\!000\ \text{K}$, corresponding to the median photon energy of 26.95 eV of our adopted spectrum. The virial temperature \citep{BarkanaLoeb2001}, also plotted in Figure~\ref{fig:virial}, grew with the DM halo mass roughly until it reached the temperature of the gas for each respective scenario, at which point a runaway gravitational collapse occurred. 

Figure~\ref{fig:radial} shows density, temperature, free-electron fraction, and H$_2$ fraction as a function of radial distance from the DM halo centre at the redshift of collapse. Once temperatures dropped to ${\sim}\,10^4\ \text{K}$, recombination occurred, allowing the gas to cool further through electronic transitions and for shielding of the ionizing portion of the spectrum.  However, even as the gas collapsed, the free-electron fraction remained high in the scenarios with the highest-intensity radiation fields (B and C), as shown in the bottom left panel of Figure~\ref{fig:radial}. The free electrons, in turn, catalysed the formation of H$_2$ through the reactions \citep{Galli1998, Glover2005}:
\[
e^{-} + \text{H} \longrightarrow \text{H}^{-} + \gamma
\]
\[
\text{H}^{-} + \text{H} \longrightarrow \text{H}_2 + e^{-}
\]
\indent Figure \ref{fig:slice} shows density and temperature slices of each scenario at the moment the simulations were stopped. The plot shows that hydrogen deuteride (HD) cooling allowed the gas to reach the temperature floor set by the cosmic microwave background ($T_{\rm CMB}$ $\sim$ 70 K) in Scenario A, although in Scenarios B and C the LW flux was sufficient to noticeably reduce the H$_2$ (and HD) fraction, precluding this efficient cooling process. 

In Figure~\ref{fig:enclosed}, we estimate the mass enclosed in each of the three scenarios. This mass is determined by computing the infall time for each cell based on its radial velocity relative to the location of the cell with the highest gas density. We assume that gas from outflowing cells is not accreted, and therefore it is not included in the infall time calculation. We find that the mass enclosed increases along with the radiation background. This is consistent with the fact that in each respective scenario, more gas is available because the DM halo has grown larger. 

The mass available to form Pop III stars or DCBHs in each scenario can be computed based on the mass that collapses within the time limit for Pop III formation. The largest stars formed would have a lifetime of around 3 Myr, after which Pop III star formation depends on the metal mixing time-scale. \citet{sarmento2025} estimated a 10-30 Myr limit for Pop III formation, and \citet{Fotopoulou2024} showed that low-metallicity star-forming clouds could be disturbed by stellar feedback and supernovae in less than 10 Myr. Therefore, we annotate 3 and 30 Myr time-scales in Figure \ref{fig:enclosed}, to show minimum and maximum time-scales for Pop III formation. In Figure \ref{fig:accretion}, we also estimate the accretion rate onto the protostars that would form, plotted against the mass enclosed, which is computed using the same method as in Figure~\ref{fig:enclosed}. The accretion rate is estimated using the relation $\dot M \sim c_{\rm s}^3/G$, where $c_{\rm s}$ is the sound speed, assuming an ideal gas, and $G$ is the universal gravitational constant. This estimation shows that the accretion rate also increases along with the radiation background, resulting in a more top-heavy IMF for the objects that form in more intense radiative environments.

\section{Discussion}
\label{Sec:discussion}
\subsection{Pop III vs DCBH Formation}
\label{Sec:dcbh}
Whether the collapsing gas eventually forms a large Pop III cluster or a DCBH depends on the IMF of the objects that form, which is in turn dependent on the accretion rate onto the collapsing cloud. 
If accretion is sustained at rates between 0.01 and 0.1 M$_\odot$ yr$^{-1}$, then a supermassive star (SMS) would likely form \citep{Inayoshi_2020}.  The lifetime of such an SMS is expected to be around 1 Myr \citep{Wise_2019}. Once the nuclear fuel is exhausted, the SMS would collapse into a DCBH with no supernova, unless the SMS has a mass in a narrow band around 55,000 M$_\odot$, in which case a powerful supernova would occur \citep{Chen2014}, leaving behind no remnant. If accretion rates exceed $\sim 0.3$ M$_\odot$ yr$^{-1}$, collapse would happen during the helium-burning stage because of a general relativistic instability \citep{Umeda_2016}. In each case, the resulting black hole is expected to have a mass of $\sim 10^5$ M$_\odot$. If accretion rates are not sustained above $\sim 0.01$ M$_{\odot}$ yr$^{-1}$ during the first $10^5$ yr (the Kelvin-Helmholtz time-scale), then a star with a mass of order $\sim 100$ M$_\odot$ would likely form, its mass limited by its own radiative feedback \citep{Haiman_2012}. 

The accretion rates are dependent on the sound speed, which is in turn dependent on the gas temperature. Therefore, the fate of our clusters can be determined through Figures~\ref{fig:enclosed} and~\ref{fig:accretion}. In Scenario A, a more bottom-heavy IMF Pop III cluster would form, because less than 100 M$_\odot$ will have been accreted after $10^5$ years (Figure \ref{fig:enclosed}), which is also consistent with the low accretion rates in Figure \ref{fig:accretion}. Scenario B would result in a more top-heavy IMF Pop III cluster, because of the higher accretion rate, although a SMS is still ruled out because the accretion rates remain below the threshold until after the Kelvin-Helmholtz time. Only in Scenario C may a DCBH form, because of the sustained high accretion rates. 

Thus, we find that for our DM halo of mass $2.23 \times 10^{9}$ M$_{\odot}$ at $z = 12.63$, with an intense background corresponding to a $10^9$ M$_\odot$ SMBH with the Eddington luminosity 10 kpc away, a DCBH is most likely to form. In Scenarios A and B, corresponding to the same SMBH at respective distances of 1000 and 100 kpc, a large cluster of Pop III stars is instead likely to form, at $z = 24.99 \rm\ and\ 15.46$, respectively, with the second cluster having a more top-heavy IMF. 
\subsection{Detectability by \textit{JWST}}
\citet{Maiolino2024} observed a Pop III candidate, identified by its He\,{\large \sc ii} $\lambda$1640 emission line, at an estimated distance of 2.5 kpc from the high-redshift galaxy GNz-11 at $z = 10.6$.  They estimated a luminosity in this line of order $10^{41}\ \mathrm{erg\ s}^{-1}$. The luminosity of the clusters in our simulations depends on the star formation efficiency and the IMF of the stars that are formed. Recent work has suggested that star formation efficiency may be elevated  at high redshifts, with estimates ranging from around 1\% to greater than 10\% \citep{Andalman2025,Venditti2024,Meidt2025,Somerville2025,Ventura2024,Bovill2024}, with \citet{Gentile2024} even suggesting that star formation efficiencies for Little Red Dots could be 25-50\%. \citet{kar2025extremeburstinessevolvingstar} also suggested that star formation efficiency could be a function of halo mass and redshift. We therefore assume a 10\% star formation efficiency, and a top heavy IMF \citep{Bromm_2001}. We also note that \citet{wasserman2025ultravioletphotonproductionrates} showed that rotating Pop III stars would have enhanced He\,{\large \sc ii} $\lambda$1640 emission, reducing the top-heaviness of the IMF that would be necessary for detection.

Using the same method as \citet{Johnson2010}, we compute an expected luminosity of our Pop III clusters in all three of our simulated scenarios. The luminosity emitted by a Pop III cluster in the He\,{\large \sc ii} $\lambda$1640 emission line can be approximated as
\begin{align}
L_{1640} \simeq\ &4 \times 10^{38} \ \mathrm{erg\ s}^{-1} \times \left( \frac{1 - f_{\mathrm{esc}}}{1} \right) \times \left( \frac{f_*}{0.01} \right) \times \left( \frac{f_{\mathrm{coll}}}{0.1} \right) \notag \\
&\times \left( \frac{Q_{\mathrm{HeII}}}{5 \times 10^{45}\ \mathrm{s}^{-1} \ \mathrm{M_\odot^{-1}}} \right) \times \left( \frac{M_{\mathrm{h}}}{10^8\ M_\odot} \right),
\end{align}
where $f_{\rm esc}$ is the photon escape fraction, $f_*$ is the star formation efficiency, $Q_{\rm HeII}$ is the number of He-II ionizing photons ($\rm M_{\odot}^{-1} \rm\ s^{-1} $) in the stars that form, $f_{\rm coll}$ is the fraction of the gas that cools and collapses, and $M_{\rm h}$ is the DM halo mass. The corresponding observed flux is:

\begin{align}
\label{eqn:flux}
F_{1640} \simeq\ 10^{-20} \ \mathrm{erg\ s}^{-1} \mathrm{\ cm}^{-2} \times \frac{L_{1640}}{10^{40} \ \mathrm{erg\ s}^{-1}} 
&\quad \times \left( \frac{1+z}{10} \right)^{-2}
\end{align}

We calculate the gas collapse fraction $f_{\rm coll}$ used in Equation \ref{eqn:flux} by dividing the gas enclosed (shown in Figure \ref{fig:enclosed}) by the gas mass within the virial radius of the DM halo.
Using $Q_{\rm HeII} = 3.8\ \times\ 10^{47}\ \rm M_{\odot}^{-1} \rm\ s^{-1}$ (based on the top-heavy IMF), an escape fraction of zero, and a star formation efficiency of 10\%, values of L$_{1640}$ and F$_{1640}$ are shown in Appendix B for the three different scenarios, at three different times post collapse. 

\begin{table}
    \centering
    \caption{
        Signal-to-noise ratios (SNRs) for three simulated scenarios using \textit{JWST}/NIRSpec. 
        SNRs for the He\,\textsc{ii} $\lambda1640$ flux are listed for cluster masses corresponding to the enclosed masses with infall times of 3, 10, and 30 Myr shown in Figure~\ref{fig:enclosed}. }
    \label{tab:snr}
    \begin{tabular}{lcccc}
        \hline
        Scenario & Collapse $z$ & SNR (3 Myr) & SNR (10 Myr) & SNR (30 Myr) \\
        \hline
        A & 24.99 & 0.06 & 0.18 & 0.59 \\
        B & 15.46 & 1.74 & 3.39 & 9.32 \\
        C & 12.63 & 3.25$^\ast$ & 14.07$^\ast$ & 46.89$^\ast$ \\
        \hline
    \end{tabular}
\raggedright \footnotesize \\ $^\ast$For Scenario C, we show in Section~\ref{Sec:dcbh} that a DCBH is likely to form. This DCBH could be accompanied by a stellar population and/or other DCBHs. The SNRs shown for Scenario C represent the maximum possible stellar contribution.
\end{table}
\citet{Maiolino2024} observed the candidate Pop III cluster near GN-z11 using the \textit{JWST} NIRSpec instrument, in integral field spectroscopy (IFS) mode. They had an exposure time of 10 hours 46 minutes, with the G235M/F170LP grating/filter pair. Using the \textit{JWST} exposure time calculator \citep{Pickering_2016}, we calculate the SNRs of our three clusters under the same observing strategy, except for Scenario A at $z = 24.99$, where we instead use the G395H/F290LP grating/filter pair (the emission line at $z = 25$ is redshifted out of the range of the G235M/F170LP pair). We list the signal-to-noise ratios (SNRs) for the fixed slit observing mode instead of the IFS mode, to simplify the calculation; we found that the IFS SNRs were comparable. These SNRs are listed in Table \ref{tab:snr}. We find that Scenario A is not detectable without the help of significant gravitational lensing. On the other hand, Scenarios B and C could be detectable.

We can make a direct comparison between GNz-11 and our model SMBH, by assuming that GNz-11 hosts a black hole of around $2 \times 10^6$ M$_\odot$, and accretes at around five times the Eddington rate \citep{maiolino2024smallvigorousblackhole}. \citet{Maiolino2024} estimated that the He\,{\large \sc ii} clump was around 2.5 kpc away from GNz-11. Therefore the flux from GN-z11 at the distance of the He\,{\large \sc ii} clump would be around 0.16 times the flux in Scenario C, and 16 times the flux in Scenario B. Assuming the mass infalling within 30 Myr is also intermediate between scenarios B and C, at around $\sim 10^7 $ M$_\odot$, and again assuming a star formation efficiency of 10\% and an escape fraction of zero, then the expected luminosity of the He\,{\large \sc ii} clump would be $\simeq$ $3\times10^{41}$ erg s$^{-1}$, which is in line with the luminosity that was actually observed. 

Therefore, we conclude that Scenarios B and C could be detectable with \textit{JWST}. Scenario A, even assuming a top-heavy IMF, when in fact the cluster would likely have a more bottom-heavy IMF than Scenarios B and C, would only have an SNR of up to 0.59, meaning that it is very unlikely to be detectable with \textit{JWST}. In Scenario B, we conclude that a large, pure Pop III cluster could be detectable with \textit{JWST}'s NIRSpec instrument.  In Scenario C, the collapsing gas will form one or more DCBHs, which may be accompanied by a detectable stellar population. 
\subsection{Time-Dependent and Inhomogeneous Effects}
AGN exhibit considerable variation in luminosity over time. Here we discuss whether this variation could cause an early collapse of the gas. For an ionized gas, the strongest interaction at high redshift would be Compton scattering between free electrons and CMB photons. If the AGN background were turned off, the temperature of the gas would decay exponentially, approaching the temperature of the CMB. A characteristic Compton cooling time \citep{Benson_2010} is calculated as \begin{equation}
t_{\rm comp} = \frac{3 m_{\rm e} c \left( 1 + \frac{1}{x_{\rm e}} \right)}{8 \sigma_{\rm t} a T_{\rm CMB}^4 \left( 1 - \frac{T_{\rm CMB}}{T_{\rm e}} \right)}
\end{equation} where $x_{\rm e} = \frac{n_{\rm e}}{n_{\rm t}}$, $n_{\rm e}$ is the electron number density, $n_{\rm t}$ is the number density of all species, $T_{\rm CMB}$ is the temperature of the CMB, $a$ is the radiation constant, and $T_{\rm e}$ is the electron temperature. Therefore, at $z = 15$, the characteristic Compton cooling time would be 35.5 Myr, meaning that any  time-dependent variability in the AGN luminosity out to this timescale would not result in an early collapse. Average quasar duty cycles are thought to be below 0.01 \citep{Yu_2002,Chen_2018, Eilers2024}. \citet{_urov_kov__2024} showed that the average quasar lifetime at $6 \leq z \leq 7$ is $\leq 7$ Myr, although quasar lifetimes and duty cycles at the high redshift studied in this article ($z > 12$) are uncertain. 

This suggests that the scenarios we studied, under the assumption of persistent radiation from a nearby quasar, are likely to be uncommon. If an AGN were only present for a fraction of the time, its heat would remain in the gas over the Compton cooling time, helping to prevent gas collapse over that period. The AGN would still act to increase the collapse time, albeit not as effectively as if the AGN were continuously emitting. The AGN lifetime and duty cycle will thus impact the resulting collapse time. Given the highly uncertain nature of black hole growth at these highest redshifts, our assumption of a 100\% duty cycle can be considered an upper limit of the effect of a true variable AGN near a halo.  

In addition to ignoring AGN temporal variability, a second limitation of our simulations is that we apply the AGN radiation as an isotropic radiation field. Previous work \citep{Visbal2017} showed that in simulations of the effect of an ionizing UV background, the time of collapse was similar when both a point source and an isotropic radiation field were used, provided that there is a correction for local extinction when using the isotropic radiation field. Our simulations also do not consider any spatial anisotropy of the AGN emission, caused for example by dust obscuration. We assume that all of the emission reaches the neighbouring DM halo, isotropically at a constant rate; dust obscuration along a specific sightline would reduce the incident quasar flux. Overall, both dust obscuration and variability would decrease the incident flux, and thus collapse could occur sooner. Our work focuses on the idealized scenario, to estimate the maximal effect of such an AGN on a nearby growing DM halo.

\section{Conclusions}
\label{Sec:conclusion}
We have presented three-dimensional radiation-hydrodynamics simulations of three scenarios, corresponding to DM haloes exposed to the radiation from an SMBH with a bolometric luminosity of $1.26 \times 10^{47}$ erg s$^{-1}$. We ran simulations with background radiation fields of the SMBH at respective distances of 1000, 100, and 10 kpc from our DM halo (Scenarios A, B, and C, respectively), using an isotropic radiation background, calculated using an SED \citep{HM2012} that extends from $3-10^6\ \rm eV$. Our main findings are the following:

\begin{itemize} 
\item The high-energy portion of the SED has a significant effect on the collapsing gas. Compton scattering maintains a high free-electron fraction even as the gas cools and collapses. This high free-electron fraction in turn catalyses H$_2$ formation, which allows further cooling of the gas through H$_2$ radiative quadrupole transitions and HD dipole transitions. Moreover, Compton scattering heats the ionized gas to higher temperatures than possible with only an ionizing UV background. This further delays gravitational collapse until the virial temperature exceeds the Compton equilibrium temperature, which allows more gas to be available for Pop III formation. In our simulations, this corresponds to collapse at $z = 24.99, 15.46$ and 12.63 for Scenarios A, B, and C respectively. These effects are not captured by previous work simulating the effect of an ionizing radiative background on Pop III formation \citep[e.g,][]{Visbal2017,Kulkarni2019,Park2021}.

\item The level of the radiation background changes the IMF of the objects formed after collapse. Scenario A corresponds to a lower-IMF Pop III cluster, Scenario B to a high-IMF Pop III cluster, and Scenario C to a DCBH, which may be accompanied by a stellar population. \citet{Maiolino2024} found that the IMF of the stars powering the He\,{\large \sc ii} clump near GNz-11 must extend to at least 500 M$_{\odot}$. This is consistent with our simulations, which show that an intense radiation background would make the IMF of a Pop III cluster more top-heavy.

\item We find that Pop III clusters formed near AGN may be detectable in He\,{\large \sc ii} $\lambda$1640 out to $z \sim 15$ using \textit{JWST}'s NIRSpec instrument, in both fixed slit and IFS modes.
\end{itemize}
Our results are consistent with recent observational work \citep{Zhu2025} which showed that a quasar, which shined for $\sim 3.1$ Myr,  suppressed recent star formation within $\sim7$ comoving Mpc. In particular, this observation reflects the relevance of our scenario C in which we would expect a DCBH to form. Our work is also consistent with \citet{Maiolino2024}, which observed possible signatures of Pop III stars near GN-z11. This observation is consistent with Scenarios A and B, where we would expect Pop III clusters to form.

\section*{Acknowledgements} 
The work presented in this article was performed at Los Alamos National Laboratory (LANL) under the auspices of the United States National Nuclear Security Administration. LANL is operated by Triad National Security LLC (Contract No. 89233218CNA000001). MAM acknowledges support by the Laboratory Directed Research and Development program of LANL under project number 20240752PRD1. The authors are thankful to an anonymous reviewer for their helpful suggestions.
\section*{Data Availability}
All of the simulations presented in this article were run using {\Large\sc enzo} v2.6.1 (commit hash 399df69). The source code was modified to add isotropic photoionization, photoheating, and photodissociation rates, and a treatment for Compton scattering off of our background. Initial conditions were generated using {\Large\sc music} using the coarse grid seed 27700. The seeds for levels 8 and 9 were 44278, and 66886, respectively. All computational analysis of the snapshots generated was done using \texttt{yt}. The source code for these three programs is publicly available. The {\Large\sc enzo} parameter files used, along with the final snapshots, are available upon request from EMF.



\bibliographystyle{mnras}
\bibliography{main} 




\appendix

\section{Computed rates}
In Table \ref{tab:photo_rates}, we show the isotropic photoionization and photoheating rates for Scenario B. The LW rate for Scenario B was $J_{21} = 5232$. 
Rates for Scenarios A and C can be computed by multiplying, or dividing, respectively, the given rates by 100.

\begin{table}
    \centering
    \caption{
        Per-particle photoionization and photoheating rates for hydrogen, neutral helium, and singly ionized helium. Rates are computed using the adopted SED and are given in units of s$^{-1}$ and erg s$^{-1}$, respectively.
    }
    \label{tab:photo_rates}
    \begin{tabular}{lccc}
        \hline
        Process & H & He & He$^+$ \\
        \hline
        Photoionization rate & $1.25 \times 10^{-8}$ & $1.08 \times 10^{-8}$ & $3.47 \times 10^{-10}$ \\
        Photoheating rate  & $8.00 \times 10^{-20}$ & $1.42 \times 10^{-19}$ & $8.93 \times 10^{-21}$ \\
        \hline
    \end{tabular}
\end{table}

\section{Detectability Calculations}
 The expected luminosities of the clusters, and the fluxes that could be observed by \textit{JWST}, are listed in Tables \ref{tab:luminosity} and \ref{tab:flux}, respectively.
 \begin{table}
    \centering
    \caption{He\,{\large \sc ii} $\lambda$1640 luminosities ($\rm erg\ s^{-1}$) for the three simulated scenarios, at three characteristic time-scales after runaway gravitational collapse. In comparison, the He\,{\large \sc ii} clump from \citet{Maiolino2024} has an estimated luminosity of $\sim10^{41}\ \mathrm{erg\ s}^{-1}$. }
    \label{tab:luminosity}
    \begin{tabular}{lccc}
        \hline
        Scenario & 3 Myr & 10 Myr & 30 Myr \\
        \hline
         A  & $1.18\times10^{40}$ & $3.55\times10^{40}$ & $1.18\times10^{41}$ \\
        
        B & $1.91\times10^{41}$ & $3.82\times10^{41}$ & $1.15\times10^{42}$ \\
        C  & $3.20\times10^{41}$ & $1.60\times10^{42}$ & $8.01\times10^{42}$ \\
       
        \hline
    \end{tabular}
\end{table}

\begin{table}
    \centering
    \caption{Observed He\,{\large \sc ii} $\lambda$1640 fluxes ($\rm erg\ s^{-1}\ cm^{-2}$) for the three simulated scenarios using \textit{JWST}/NIRSpec.}
    \label{tab:flux}
    \begin{tabular}{lccc}
        \hline
        Scenario & 3 Myr & 10 Myr & 30 Myr \\
        \hline
        A  & $1.75\times10^{-21}$ & $7.01\times10^{-21}$ & $1.75\times10^{-20}$ \\
        B & $7.04\times10^{-20}$ & $1.41\times10^{-19}$ & $4.23\times10^{-19}$ \\
        C  & $1.73\times10^{-19}$ & $8.66\times10^{-19}$ & $4.33\times10^{-18}
$ \\
        \hline
    \end{tabular}
\end{table}


\bsp	
\label{lastpage}
\end{document}